\documentclass{ws-mpla}
\usepackage[super]{cite}
\usepackage{graphicx}
\begin{document}

\markboth{Peng et al.,} {A Magnetic-Monopole-Based Mechanism to the
formation of the Hot Big Bang Modeled Universe}

\catchline{}{}{}{}{}

\title{A Magnetic-Monopole-Based Mechanism to the formation of the Hot Big Bang Modeled Universe
\footnote{This work was supported in part by the National Natural
Science Foundation of China under grants 11565020, 11965010, and the
Counterpart Foundation of Sanya under grant 2016PT43, 2019PT76 the
Special Foundation of Science and Technology Cooperation for
Advanced Academy and Regional of Sanya under grant 2016YD28, the
Scientific Research Starting Foundation for 515 Talented Project of
Hainan Tropical Ocean University under grant RHDRC201701, and the
Natural Science Foundation of Hainan Province of China under grant
118MS071, 2019RC239.} }

\author{\footnotesize Qiu-He Peng$^{1}$ and Jing-Jing Liu$^{2}$\footnote{
Corresponding author.}}

\address{$^{1}$Department of Astronomy, Nanjing University,
Nanjing, Jiangshu 210000, China\\
$^{2}$College of Marine Science and Technology, Hainan Tropical
Ocean University, Sanya, Hainan 572022, China\\
liujingjing68@126.com}

\author{Chi-Kang Chou$^{3}$}

\address{$^{3}$National
Astronomical Observatory, Chinese Academy of Sciences, Beijing,
100000, China}

\maketitle

\pub{Received (Day Month Year)}{Revised (Day Month Year)}

\begin{abstract}
There are some particle physics theories that go beyond the
so-called ''standard cosmological model'' to predict the existence
of magnetic monopoles\,(MMs). The discovery of magnetic monopoles
would be an incredible breakthrough in high-energy physics. The
existence of MMs in the early Universe has been speculated and
anticipated from Grand Unified Theory. If MMs exist, the inverse
powers of the unification mass will not suppressed the baryon number
violating effects of grand unified gauge theories. Therefore, MM
catalyzing nucleon decay is a typical strong interaction. This
phenomenon is due to the boundary conditions that must be imposed on
the core of MM fermion fields. We present a possible mechanism to
explain the formation of the Hot Big Bang Cosmology. The main
ingredient in our model is nucleon decay catalyzed by magnetic
monopoles (i.e., the Rubakov-Callan effect). It is shown that Hot
Big Bang developed naturally, because the luminosity due to the
Rubakov-Callan effect is much greater than the Eddington luminosity
(i.e., $L_m>10^4L_{\rm{Edd}}$).

\end{abstract}

\ccode{PACS Nos.: 97.60.Bw, 26.30.Jk, 23.40.-s.}

\section {Introduction}
On the standard model of the Hot Big Bang Cosmology, the early
Universe is depicted by extrapolating back to a hot and dense
initial state of Planck length and Planck time derived according to
the uncertainty principle. In 1992, the origin of the Big Bang was
discussed by Thakur(1992)\cite{Thakur1992}. They developed a
singularity-free model of the universe within the framework of the
Friedmann-Lemaitre-Robertson-Walker cosmology. Then the Big Bang
model on its origin and development have been investigated by
Alpher(1999)\cite{Alpher1999}. The standard cosmological model and
some related issues have been discussed by some astronomers and
scholars(e.g., \cite{Mercik2006, Jain2010, Shaposhnikov2010,
Bellini2012, Figueroa2017}). Recently, a macroscopic view of the
standard cosmological model was given by Ignat'ev et al.,
(2018)\cite{Ignat'ev2018}. Khlopov (2018) also discussed the
standard models of particle physics and cosmology
\cite{Khlopov2018}. Their results showed that the modern Standard
cosmological model of inflationary Universe and baryosynthesis is
deeply involved particle theory beyond the Standard model.
Brian.(2018), also redefined the standard model of
cosmology\cite{Brian2018}. However, the formation of the Hot Big
Bang itself has not been investigated. In this paper, we will
present a possible mechanism to explain the formation of the Hot Big
Bang Cosmology. The main ingredient in our model is nucleon decay
catalyzed by magnetic monopoles\,(MMs).

The experiment tells us that the magnetic north and south poles
cannot be divided into magnetic monopoles\,(MMs), i.e., isolated
magnetic charges. Petrus (1269) discussed this issue in the 13-th
century. There are strong theoretical reasons to believe that MMs
should exist\cite{Peregrinus1269}. Hooft and Polyakov (1974) showed
that MMs are an inevitable prediction of Grand Unified Theory of
elementary particle interactions\cite{Hooft1974, Polyakov1974}, and
the same is generally true for more modern theories of everything,
such as superstring theory\cite{Duff1995}. Therefore, it would be an
incredible breakthrough if people could find a MM particle in
high-energy physics.

Although magnetic monopole has not been discovered, it plays an important role in the theoretical research of high energy physics.For example, magnetic monopoles provide powerful theoretical tools for studying properties of strongly coupled non-abelian canonical field theory, such as quantum chromodynamics, and in particular its supersymmetric variants

Although MMs have not been found, they play an important role in
theoretical research of high-energy physics. For example, the MMs
give powerful theoretical tools for exploring properties of strongly
coupled non-Abelian gauge field theory, such as quantum
chromodynamics, and in particular its supersymmetric
variants\cite{Shifman2007}. Therefore, the study MMs and its related
problems\,(e.g., search for MMs) is a hot topic in the field of
high-energy physics and astrophysics
\cite{Abbasi2013,Vandewalle2014, Fujii2015, Aab2016, Pollmann2018,
Bazeia2018, Peng2018, Nakosai2019,  Frank2019}.

The existence of MMs in the early Universe has been speculated and
anticipated from grand unification. The goal of such theories is to
unity the strong, weak, and electromagnetic interaction in terms of
quarks and leptons within the framework of a gauge field theory
based on non-Abelian symmetry group. These theories make two
startling predictions: namely, the instability of the proton, and
the existence of stable and heavy MMs.

Most of physicists believe that the existence of MMs had been ruled
out by experiments. However, experiments only indicated that the
flux of MMs on the Earth is too low to be observed. We summarize
some predictions from the model of supermassive object with MMs,
which match up with recent astronomical observations quantitatively.
They may signal the presence of MMs in supermassive objects, such as
one at the Galactic Center. In 2001, we discussed ultra-high-energy
cosmic rays from supermassive objects with MMs, as well as
high-energy radiation from quasars, active galactic nuclei, and the
Galaic Center with MMs \cite{Peng2001a, Peng2001b}. Very recently,
some issue on MMs were illustrated by our group\,(e.g.,
\cite{Peng2017a, Peng2017b, Peng2018}. We discussed a series of
important but puzzling physical mechanisms concerning the energy
source, various kinds of core-collapse supernovae explosion
mechanisms during central gravitational collapse in astrophysics.
The puzzles of possible association of $\gamma$-ray burst with
gravitational wave perturbation, the heat source for the molten
interior of the core of the Earth and  the cooling of white dwarfs
were also investigated. We have made use of the estimations for the
space flux of MMs and nucleon decay induced by MMs, called the
Rubakov-Callen (RC) effect\,\cite{Rubakov1981, Rubakov1988,
Callan1983}, to obtain the luminosity due to the RC effect and also
investigated other problems related to supernova explosion (e.g.,
\cite{Liu2013, Liu2014, Liu2016a, Liu2016b, Liu2016c, Liu2017a,
Liu2017b, Liu2018a, Liu2018b, Liu2018c, Gao2014, Gao2015, Gao2014,
Gao2017a}).

Rubakov (1981, 1988) and Callan(1983) \cite{Rubakov1981,
Rubakov1988, Callan1983} have shown that the fermion wave function
is literally sucked into the core because of the potential between
the $s$-wave of a fermion and that of a MM. Due to sucking of the
$s$-wave, the catalysis cross section saturates the uncertainty
bound: $\langle \sigma v\rangle\approx E_F^{-2}$, where $E_F$ is the
Fermi energy. The actual catalysis section depends on the Grand
Unified Theory. In SU(5), we know $M+m\rightarrow M+\pi^-+e^+$ or
$M+p\rightarrow M+\pi^0+e^+$ . It is then expected that MM catalysis
has great potential to produce astrophysical fireworks, and
applications of the $R-C$ effect to quasars and active galactic
nuclei have also made remarkable achievements \cite{Peng2001a,
Peng2001b, Peng2016, Peng2017a}.

In this paper, we will use the R-C effect to explain the formation
of the Hot Big Bang. We want to propose a possible mechanism to
describe how the Big Bang  developed. The main ingredient in our
model is the MMs catalyzing nucleon decay with strong cross section
of interaction.
\section[]{The Astronomical evidence for both the absence of black holes and the existence of MMs at our Galactic Center }

Eatough et al.(2003) reported a measurement of a strong magnetic
field around the supermassive black hole at the centre of the
Galaxy\cite{Eatough2013}. We know that at $r=0.12$ pc near the
Galactic center (hereafter GC), the minimum value of outward radial
magnetic field is $B > 8$\,mG, which is larger than the Alphen
critical value $B_{\rm{Alphen}} \sim1.3$ mG. Due to the exists of
the strong radial magnetic field, the accretion (plasma) disk is
prevented from approaching to the GC, and may not enter in the
neighborhood of the GC. As a result, the radiations observed from
the accretion disk gas around the black hole of the GC are hardly to
emit.  It is a difficult situation of the standard accretion disk
model of black holes in the GC \cite{Peng2016, Peng2017a}.

Considering the RC effect,the MMs may catalyze nucleon decay, which
can be used as an energy source. However, this dilemma in the GC may
be naturally solved by our super-massive star model with
\textbf{MMs}. At $r=0.12$ pc near the GC, the observed outward
radial magnetic field strength (i.e., $B>8$\,mG) \cite{Eatough2013}
\textbf{is} in good agreement with our theoretical predictions
\cite{Peng2001a}.

In our model, at least three predictions were quantitatively
confirmed by sbusequennt observations. Firstly, the GC emits large
numbers of positrons at a rate about $10^{43}e^+/$sec or so. This is
consistent with the high-energy astrophysical observations
\cite{Kn2003}. Secondly, at $r=0.12$ pc of the super-massive object
core, the radial magnetic field strength produced by the
\textbf{MMs} condensed is about $B\approx(10\sim50)$ mG, which is
consistent with the lower limit of the magnetic field strength
observed \cite{Eatough2013}. Finally, for the super-massive stellar
object at the GC, we predicted their surface temperatures to be
about 123 K, corresponding to a frequency of  $10^{13}$ Hz (at the
sub-millimeter range). This frequncy predicted is quite close to the
observed value (i.e., $10^{12}$ Hz) \cite{Falcke2013}.

The implications of the facts that these predictions were
quantitatively confirmed by astronomical observations are: 1) There
is no supermassive black hole at the center of our galaxy; 2) These
are the evidences for the existence of MMs \cite{Peng2016,
Peng2017a}. We would like to declare that the astronomical
observations are really the physical experiments in cosmic space,
although MMs have not been detected by the physicists up to now.

\section[]{An united model of supernova explosion driven by MMs}

Regarding the RC effect as a source of energy, MMs can catalyze the
decay of nucleons. Therefore we propose a unified model supernova
explosions caused by MMs \cite{Peng2017b}. The main idea is as
follows. Taking the nucleon decay induced by MMs in particle physics
and making estimation for the space flux of MMs, we can obtain a
formula to estimate the luminosity due to the RC effect.

According to the RC luminosity formula, we {discuss the issue that
the supernova explosion would develop just when its luminosity is
much greater than the Eddington's luminosity due to RC effect in the
star. We present a unified treatment for all kinds of core-collapse
supernovae (e.g., SNII, SNIb, SNIc, Ultra-luminous supernova
(ULSN)). We will also discuss the gamma-ray burst generation
mechanism in detail in our paper. Those weak or/and dark explosions
of supernova will be also naturally expressed by our idea due to the
fact that its RC luminosity is just greater than the Eddington's
luminosity of the star.

The very important result is that no matter how massive the
supernova is, a neutron star\,(NS) will be formed after supernova
explosion due to the $R-C$ effect by a small amount MMs remained in
the new born NS. However, this theory does not apply to black holes.

Besides, both the heat source for the core of the Earth and the
energy source needed for the white dwarf interior are also solved by
the same treatment using the RC effect. The possible association of
the short gamma-ray burst detected by the Fermi gamma-ray Burst
Monitoring Satellite (GBM) and the LIGO gravitational wave event
(GW150914) may be reasonably explained by our unified model. In the
present paper, based on the same idea, we propose further a model of
hot big bang cosmology driven by MMs without initial singularity of
the Universe .

\section[]{The standard cosmology}

A most fundamental feature of the standard cosmology is the
expansion of the Universe. The expansion, discovered in the 1920's.
led to Hubble's law and played a fundamental role in observational
cosmology. Almost all the galaxy spectra measured by observers in
the world are redshifted, illustrating the universality of the
expansion with \textbf{redshift} $z$ . For a smaller distance, we
can also interpret it as a simple Doppler effect by the redshift
$z$.  Thus, we have
\begin{equation}
z=\frac{\lambda-\lambda_0}{\lambda_0}=\frac{v}{c}, \label{eq.1}
\end{equation}
where the speed of light is much larger than the secession velocity
of galaxies. $\lambda_0$ and $\lambda$ are the original radiation
wavelength and the observed wavelength, respectively. However, for a
larger distance, Eq.(1) must be replaced by
\begin{equation}
1+z=\frac{\sqrt{1+v/c}}{\sqrt{1-v/c}}. \label{eq.2}
\end{equation}

For large recession velocities of the distant galaxies, the Hubble's law
may be written as
\begin{equation}
v=Hd. \label{eq.3}
\end{equation}

In a homogeneous isotropic Universe, the expansion rate is constant
in time and the recession time of the distant galaxies may be
obtained as
\begin{equation}
t=\frac{d}{v}=\frac{1}{H_0}\approx 138~~(billion~ year),
\label{eq.4}
\end{equation}
where $0.4\leq h\leq1$, $H_0\approx65\rm{kms}^{-1}\rm{Mpc}^{-1}$,
and $H_0^{-1}$ is the age of the Universe. The redshift of galaxies
varies linearly with the distance for $z<<1$. In addition, the
homogeneous cosmology predicts the existence of the black-body
cosmic microwave background radiation and the abundances of helium
and the primordial nucleosynthesis of other elements. These are the
great achievements of the Lema\^{\i}tre model \cite{Lema1927} and
the pioneering theories of the Hot Big Bang \cite{Gamow1940,
Gamow1946, Gamow1948, Gamow1956, Alpher48, Alpher1948, Alpher1949,
Alpher1953}.

Now we describe briefly some of the relevant physics for Universe
expansion. The radius of the expanding Universe $R$ satisfies
$dR(t)/dt>0$. For homogeneous universe expansion, we have $\rho
R^3=$constant, and $T R=$constant, where $\rho$ denotes the matter
density, $T$ is the background temperature. The radiation energy
density is denoted by $aT^4$, where $a$ is the Stephen Boltzmann
constant, and the rest mass energy density is denoted by $\rho c^2$,
then the ratio is written as
\begin{equation}
\Gamma=\frac{aT^4}{\rho c^2}. \label{eq.5}
\end{equation}

We note that the early Universe could be dominated by matter or
radiation. The matter in the Universe consists of both ordinary
visible matter that can be detected and dark matter that is
invisible. The radiation is mainly comes from the cosmic background
microwave radiation. For convenience, we present the following
relevant data at the present time\cite{Weinberg2008}
\begin{equation}
\rho_M^{0}=(0.3-0.4)\rho_c=(03.-0.4)\times10^{-29}~~\rm{g~cm^{-3}},
\label{eq.6}
\end{equation}
\begin{equation}
E_M^{0}=\rho_M^{0}c^2=(0.27-0.36)\times10^{-8}~~\rm{erg~cm^{-3}},
\label{eq.7}
\end{equation}
\begin{equation}
E_r=aT_R^4,~T_R^0=2.7K,~a=7.56\times10^{-15}~~\rm{erg~cm^{-3}
K^{-4}}, \label{eq.8}
\end{equation}
\begin{equation}
E_r^0=4.0\times10^{-13}~~\rm{erg~cm^{-3}},~E_r^0<<E_M^0.
\label{eq.9}
\end{equation}
where the superscript (0) represents the present moment,
$\rho_M^{0}$ is the average density of matter in the Universe at the
present moment, $\rho_c$ is the critical density of the
corresponding material at the boundary between the closed Universe
model and the open Universe model, $E_M^{0}$ represents the average
energy density of matter in the Universe at the present moment,
$E_r$ represents the energy density of the radiation field in the
Universe, $T_R^0$ represents the temperature of the cosmic radiation
field (the cosmic background), and $a$ is the radiation constant.

We note that both the radiation energy density and the matter
density decrease with the time during the universe expansion. At the
same time, radiation from galaxy spectra are all \textbf{redshifted}
during the expansion. The density of radiant energy decays much
faster with time than the density of matter, and then radiation is
dominated in the early Universe.} About several thousand years after
the creation of the Universe, it was the dividing line between the
radiation-dominated and material-dominated ages. At the dividing
line, we have $\Gamma=1, aT^4_r=\rho_mc^2$.

Our whole universe was once in a very small volume, higher material
density and temperature. When the temperature is higher than
$10^{12}$ K or more, the universe was produced by Hot Big Bang. This
is from astronomical observations (now the universe is expanding),
we obtain that the early universe must be in the hot and dense
state. However, people don't know realistically the physical reason
of early universe by hot big bang (i.e., the outbreak mechanism of
early universe). The physical reason of the Hot Big Bang theory of
the Universe has never been discussed up to now really.

There are many speculations about the Hot Big Bang. For example,
when the time extrapolates from the present moment back to the
initial singularity ($t=0$), we have $R\rightarrow 0,
\rho\rightarrow\infty , T\rightarrow \infty$, so the Universe will
arise singularity. These are very interesting theoretical
speculations such as the baby Universe, and the Universe wave
function proposed by Hawking based on the uncertainty principle, but
the physical reason for the formation of the Hot Big Bang itself has
never been investigated. Using the $R-C$ effect we will propose a
possible mechanism for the Hot Big Bang in the next section.
\section[]{An oscillating model of the Universe}
In history, some oscillating Universe models had been proposed, but
the physical reason of the Universe expansion during the oscillation
has not been discussed. The continuum and continuum-particle models
for both oscillating and ever-expanding model universes are
considered, with the consequences of various interactions between
the components being traced and the connection with the
irreversibility of oscillating models noted by Landsberg \&
Reeves\cite{Landsberg1982}. Similar to the models above, our model
of the Universe is also the oscillating model between the expansion
phase due to the Hot Big Bang and the contracting phase due to the
gravitational attraction, but we propose the physical mechanism for
the Hot Big Bang of the Universe in this paper.

At the present time $t_0$, the radius of the Universe is $R_0$,
baryon density $\rho_0$, and the cosmic background temperature $T$,
then at a later time $t$ during the contracting phase, the
corresponding radius of the Universe $R$, the baryon density $\rho$,
and the cosmic background temperature  $T$ are given by
\begin{equation}
\frac{R}{R_0}=(\frac{\rho}{\rho_0})^{-1/3}, \label{eq.10}
\end{equation}
\begin{equation}
\frac{T}{T_0}=(\frac{R}{R_0})^{-1}=(\frac{\rho}{\rho_0})^{1/3},.
\label{eq.11}
\end{equation}

It is generally estimated and believed that there are
$2\times10^{11}$ galaxies. Every galaxy is roughly the size of over
Milky galaxy with $10^{11}$ stars, then the total number of stars in
the Universe is about $2\times10^{22}-4\times10^{22}$. The mass of
the sun is $2\times10^{33}$g, then the total mass of the Universe of
the baryons is $2.0\times10^{56}$g and the total number of the
baryons is $10^{80}$. If the content of the magnetic monopoles of
the same polarity contained in the Universe is $\zeta\equiv
N_m/N_B=10^{-20}(\zeta/\zeta^{\rm{up}})$, here $N_m$ and $N_B$ are
the number of magnetic monopole and baryons, respectively. Here
$\zeta^{\rm{up}}$ is the Parker upper limit $\zeta^{\rm{up}}\approx
10^{-20}$\cite{Parker1970}. So the total number of the magnetic
monopoles of the same polarity contained in the Universe may be
estimated to be $N_m=10^{61}(\zeta/\zeta^{\rm{up}})$. The magnetic
monopoles in the high temperature baryon plasma are strongly
compressed and moving very fast toward the center via
electromagnetic interaction.

The $R-C$ luminosity produced due to catalyzing nucleon decay by the
MMs is given \cite{Peng2001a, Peng2017a} as
\begin{eqnarray}
 L_{\rm{M}}&=\frac{4\pi}{3}r_c^3n_mn_B\langle \sigma v_T\rangle m_Bc^2=N_mn_B\langle \sigma v_T\rangle m_Bc^2,\nonumber\\
 &=10^{75}(\frac{\overline{n_B}}{n_{\rm{nuc}}})(\frac{\zeta}{\zeta^{\rm{up}}})(\frac{\sigma
 (\rm{RC})}{10^{-30}\rm{cm}^2})~~~\rm{ergs/s}.
\label{eq.12}
\end{eqnarray}

When the total mass of the Universe is compressed into supermassive
body, the corresponding Eddington luminosity is given by
\begin{equation}
 L_{\rm{Edd}}=10^{38}(\frac{M}{M_{\bigodot}})\approx 10^{61}~~\rm{ergs/s}
\label{eq.13}
\end{equation}

If the whole Universe is compressed such that
\begin{equation}
(\frac{\overline{n_B}}{n_{\rm{nuc}}})(\frac{\zeta}{\zeta^{\rm{up}}})(\frac{\sigma
 (\rm{RC})}{10^{-30}\rm{cm}^2})>10^{-10},
\label{eq.14}
\end{equation}
i.e.,
\begin{equation}
(\frac{\overline{n_B}}{n_{\rm{nuc}}})>10^{-10}[(\frac{\zeta}{\zeta^{\rm{up}}})(\frac{\sigma
 (\rm{RC})}{10^{-30}\rm{cm}^2})]^{-1}, \label{eq.15}
\end{equation}
Then $L_m>10^4L_{\rm{Edd}}$ and the whole Universe must violently
explode outward leading naturally to the Hot Big Bang. From Eq.(10), we my estimate the radius of the Universe at the Hot Big Bang to be
\begin{equation}
R\approx3\times10^{-12}R_0[(\frac{\zeta}{\zeta^{\rm{up}}})(\frac{\sigma
 (\rm{RC})}{10^{-30}\rm{cm}^2})]^{1/3}, \label{eq.16}
\end{equation}
i.e.,
\begin{equation}
R\approx3\times10^{-2}[(\frac{\zeta}{\zeta^{\rm{up}}})(\frac{\sigma
 (\rm{RC})}{10^{-30}\rm{cm}^2})]^{1/3}~~~\rm{pc}, \label{eq.17}
\end{equation}
where we have used the present radius of the Universe
$R_0\approx10^{10}$pc and the present average baryon density given
from Eq.(6). The temperature at the Hot Big Bang also can be
estimated to be
\begin{equation}
\frac{T}{T_0}\approx3\times10^{11}[(\frac{\zeta}{\zeta^{\rm{up}}})(\frac{\sigma
 (\rm{RC})}{10^{-30}\rm{cm}^2})]^{-1/3}, \label{eq.18}
\end{equation}
where $T_0 =2.7$\,K.

In the traditional standard Hot Big Bang Cosmology, it is
extrapolated back to the initial singularity of the Universe. This
is done purely by theoretical speculation. Our model of the Hot Big
Bang is obtained in terms of the Rubakov-Callan luminosity and no
other theoretical arguments or anticipations are required. In our
model, the expansion phase may finally end, followed by the
contraction phase due to gravitational attraction.

\section{Summary and Discussions}
In this paper, we have used the $R-C$ effect to explain the
formation of the Hot Big Bang and presented a possible mechanism
that can delineate the details of how the Big Bang developed. The
main ingredient in our description of the Hot Big Bang is MMs
catalyzing nucleon decay with strong interaction cross section. Our
results showed that whether the Universe is in an accelerating
expansion phase needs further discussion. On the other hand, the
direct observational evidence on the dark energy is also lost by the
observational error analyses of SNIa. Our model of the Hot Big Bang
is obtained in terms of the Rubakov-Callan luminosity and no other
theoretical arguments or anticipation are required. In our model,
the expansion phase may finally end followed by the contraction
phase due to gravitational attraction.

The popular view of indirect observational evidence for the
accelerating expansion of the universe comes from the comparison of
theoretical simulations of the accelerating expansion of the
universe and the deviation observation of the isotropy of the cosmic
microwave background temperature using WMAP satellite observation
data. The popular idea of indirect observational evidence for the
accelerating expansion of the Universe comes from the comparison of
theoretical simulations of the Universe accelerating expansion and
the Universe with the observational data of WMAP satellite for the
deviation observation of the isotropy of the cosmic microwave
background temperature using WMAP satellite observation data.
 In recent years, the results of some research groups are in line with
our ideas. For instance, Nielsen et al. (2016) analyzed recent
observations of a group SNIa\,\cite{Nielsen2016}. Their conclusions
did not support the accelerating expansion of the Universe. More
recently, David et al. (2017) also did not support the idea of the
accelerating expansion of the Universe\cite{David2017}.

As is well known, the 2011 Nobel prize awarded to three astronomers
(i.e., Riess, Schmist and Perlmutter), because they proposed the
accelerating expansion of the Universe, which was confirmed by
astronomical observations.  The method of this paper is based on Guy
et al. (2007). All these researches are based on the SNIa standard
candle assumption. In fact, although the average error provided by
the UNION2 was only 0.16\,m because 685 SNIa completed big samples.
We used the observational data of the 685 SNIa from UNION2 to
reexamine and analyze the average error and found that the average
total observational error of SNIa is obviously greater than
$0.55^m$, so we can not decide whether the Universe is accelerating
expansion or not\,\cite{Peng2014, Peng2016}. In our oscillating
model of the Universe, the expanding matter is gradually
decelerating due to the Newtonian attraction during the Universe
expansion process. When the kinetic energy of the expanding matter
is lower than the potential energy of the whole Universe relative to
the expanding matter, the Universe expansion contracts.

\section*{Acknowledgments}

We would like to thank Prof. Daniel Wang, Prof. Y.F. Huang, Prof.
P.F. Chen and Prof. J. L. Han for their help to inform us some new
information of observations. This work was supported in part by the
National Natural Science Foundation of China under grants 11565020,
11965010, and the Counterpart Foundation of Sanya under grant
2016PT43, and 2019PT76, the Special Foundation of Science and
Technology Cooperation for Advanced Academy and Regional of Sanya
under grant 2016YD28, the Scientific Research Starting Foundation
for 515 Talented Project of Hainan Tropical Ocean University under
grant RHDRC201701, and the Natural Science Foundation of Hainan
Province of China under grant 118MS071.


\begin{thebibliography}{0}
\bibitem{Thakur1992} R. K. Thakur., Ap\&SS., \textbf{190}, 281(1992)
\bibitem{Alpher1999} R. A. Alpher., Odessa Astronomical Publications, \textbf{12}, 10(1999)
\bibitem{Mercik2006} A. Mercik, S. Mercik., Physics, \textbf{04}, 4024(2006)
\bibitem{Jain2010} P. Jain., S. Mitra., MPLA, \textbf{25}, 167(2010)
\bibitem{Shaposhnikov2010} M. Shaposhnikov, PPN, \textbf{41}, 862(2010)
\bibitem{Bellini2012} M. Bellini, PhLB, \textbf{709}, 309(2012)
\bibitem{Figueroa2017} D. G. Figueroa., C. T. Byrnes., PhLB, \textbf{767}, 272(2017)

\bibitem{Ignat'ev2018}Y. G. Ignat'ev, D. Y. Ignatyev., A. R. Samigullina., GrCo., \textbf{24}, 1489(2018)
\bibitem{Khlopov2018} M. Y. Khlopov., , arXiv181109222(2018)

\bibitem{Brian2018}A. R. Brian., Redefining Standard Model Cosmology, 2018, Published by Intech Open,
London, United Kingdom, ISBN: 978-1-83880-864-8

\bibitem{Peregrinus1269} P. Peregrinus., The Letter Of Petrus Peregrinus On The Magnet, A.D. 1269.
\bibitem{Hooft1974} G. 't Hooft., Nucl.Phys. B., \textbf{79}, 276(1974)
\bibitem{Polyakov1974}A. M. Polyakov., JETP Lett., \textbf{20}, 194(1974)
\bibitem{Duff1995}M. Duff., R. R. Khuri., \& J. Lu., Phys.Rept. \textbf{259}, 213(1995)
\bibitem{Shifman2007}M. Shifman., \& A. Yung., Rev.Mod.Phys. \textbf{79}, 1139(2007)
\bibitem{Abbasi2013}R. Abbasi., Y. Abdou., M. Ackermann., et al., PhRvD, \textbf{87}, 2001(2013)
\bibitem{Vandewalle2014}N. Vandewalle., S. Dorbolo., NJPh., \textbf{16}, 3050(2014)
\bibitem{Fujii2015}T. Fujii., A. C.Pierre., ICRC., \textbf{34}, 319(2015)
\bibitem{Aab2016}A. Aab., P. Abreu., M. Aglietta, PhRvD., \textbf{94}, 2002(2016)
\bibitem{Pollmann2018}A. Pollmann., EPJWC., \textbf{168}, 04010(2018)
\bibitem{Bazeia2018}D. Bazeia., M. A. Marques., R. Menezes., PhRvD., \textbf{97}, 5024(2018)
\bibitem{Peng2018}Q. H. Peng., J.J. Liu., Chih-Kang. Chou., IAUS., \textbf{337}, 390(2018)
\bibitem{Nakosai2019}S. Nakosai., S. Onoda., JPSJ., \textbf{88}, 3701(2019)
\bibitem{Frank2019}M. Frank., A. Antoshkin., C. Dukes.,  et al., E.ICRC., \textbf{36}, 888(2019)
\bibitem{Peng2001a} Q. H. Peng, C. K. Chou, ASPC, \textbf{241}, 133(2001a)
\bibitem{Peng2001b} Q. H. Peng, C. K. Chou, ApJ, \textbf{551}, 23(2001b)
\bibitem{Peng2017a} Q. H. Peng, J. J. Liu, Z. Q. Ma, NewA., \textbf{57}, 59(2017a)
\bibitem{Peng2017b} Q. H. Peng, J. J. Liu, C. K. Chou, Ap\&SS., \textbf{362}, 222(2017b)

\bibitem{Peng2014} Qiuhe Peng,Yiming Hu, Kun Wang, et al., Journal of Astrophysics and Astronomy, \textbf{35}, 253(2014)
\bibitem{Liu2013} J. J. Liu,  MNRAS., \textbf{433}, 1108(2013)
\bibitem{Liu2014} J. J. Liu, MNRAS., \textbf{438}, 390(2014)
\bibitem{Liu2016a} J. J. Liu, et al., ApJS, \textbf{224}, 29(2016a)
\bibitem{Liu2016b} J. J. Liu, et al., RAA, \textbf{16}, 83(2016b)
\bibitem{Liu2016c} J. J. Liu, et al., RAA, \textbf{16}, 174(2016c)
\bibitem{Liu2017a} J. J. Liu, et al., RAA, \textbf{17}, 107, eprint arXiv:1707.03504(2017a)
\bibitem{Liu2017b} J. J. Liu, et al., ChPhC, \textbf{41}, 510(2017b)
\bibitem{Liu2018a} J. J. Liu, et al., RAA, \textbf{18}, 8, eprint arXiv:1711.01955(2018a)
\bibitem{Liu2018b} J. J. Liu, et al., EPJC, \textbf{78}, 84(2018b)
\bibitem{Liu2018c} J. J. Liu, et al., Ap\&SS., \textbf{363}, 185(2018c)

\bibitem{Gao2014} Z. F. Gao., X. J. Zhao., D. L. Song., N. Wang., Astron. Nachr., \textbf{335}, 653(2014)
\bibitem{Gao2015} Z. F. Gao., N. Wang., Y. Xu., et al., Astron. Nachr., \textbf{336}, 866(2015)
\bibitem{Gao2017a} Z. F. Gao., N. Wang , H. Shan., X. D. Li., et al., ApJ, \textbf{849}, 19(2017a)

\bibitem{Rubakov1981} V. A. Rubakov, Soviet Journal of Experimental and Theoretical Physics Letters., \textbf{33}, 644(1981)
\bibitem{Rubakov1988} V. A. Rubakov, Reports on Progress in Physics., \textbf{51}, 189(1988)
\bibitem{Callan1983} C. Callan, Nucl. Phys. B., \textbf{212}, 391(1983)
\bibitem{Peng2001a} Q. H. Peng, C. K. Chou, Astrophys. J., \textbf{551}, 23(2001a)
\bibitem{Peng2001b} Q. H. Peng, C. K. Chou, ASPC., \textbf{241}, 133(2001b)
\bibitem{Peng2016} Q. H. Peng, J. J. Liu, C. K. Chou, Ap\&SS., \textbf{361}, 388(2016)
\bibitem{Eatough2013}R. P. Eatough, H. Falcke, R.Karuppusamy, Natrue, 501, 391( 2013)
\bibitem{Kn2003} J. Kn$\ddot{o}$dlseder, V. Lonjou, P.Jean, et al., A\&A., \textbf{411}, 457(2003)
\bibitem{Falcke2013}H. Falcke, S. B. Markoff, CQGra., \textbf{30}, 4003(2013)
\bibitem{Lema1927}G. Lema\^{\i}tre., ASSB., \textbf{47}, 49L(1927)
\bibitem{Gamow1940}  G. Gamow,\& M., F. R., The Scientific Monthly., \textbf{51}, 373(1940)
\bibitem{Gamow1946} G. Gamow, Physical Review., \textbf{70}, 572(1946)
\bibitem{Gamow1948} G. Gamow, Nature., \textbf{162}, 680(1948)
\bibitem{Gamow1956} G. Gamow, Vistas in Astronomy., \textbf{2}, 1726(1956)
\bibitem{Alpher48} R. A. Alpher, R. Herman, Nature., \textbf{162}, 774(1948)
\bibitem{Alpher1948}R. A. Alpher, R. Herman, Gamow, G. A., Physical Review., \textbf{74}, 1198(1948)
\bibitem{Alpher1949}R. A. Alpher, R. C. Herman, Physical Review., \textbf{75}, 1089(1949)
\bibitem{Alpher1953} R. A. Alpher, J. W. Follin, R, C. Herman, Physical Review., \textbf{92}, 1347(1953)
\bibitem{Landsberg1982} P. T. Landsberg, G. A. Reeves, , Astrophys. J., \textbf{262}, 432(1982)
\bibitem{Parker1970} E. N. Parker, Astrophys. J., \textbf{160}, 383(1970)
\bibitem{Weinberg2008} S. Weinberg, J. Silk, \emph{BOOK REVIEW: Cosmology}, Classical and Quantum Gravity, (Oxford University Press, 2008)
\bibitem{Nielsen2016}J. T. Nielsen., A.Guffanti., S.Sarkar., Nature Scientific Reports, 6, 35596(2016)
\bibitem{David2017} C. David F., Open Astronomy., \textbf{26}, 111(2017)
\bibitem{Peng2014} Q. H. Peng, Y. M. Hu, K. Wang., et al.,JApA, \textbf{35}, 253(2014)
\bibitem{Peng2016} Q. H. Peng, J. Zhang, Z. Q. Luo, Y. Liang, JPhCS., \textbf{665}, 2073(2016)

\end{thebibliography}
\end{document}